\documentclass[letterpaper,twocolumn,10pt]{article}
\usepackage{usenix_2020}

\usepackage{amsmath,amsfonts}
\usepackage[ruled,lined,linesnumbered]{algorithm2e}
\usepackage{graphicx}
\usepackage{xurl} 
\usepackage{xspace}
\usepackage{multirow}
\usepackage{booktabs}
\usepackage{pifont}
\usepackage{balance}

\AtBeginDocument{}

\newcommand{\para}[1]{\noindent \textbf{#1 }}

\newcommand{\sys}{AuroraRL\xspace}

\begin{document}

\title{AuroraRL: Fast, Fault-Tolerant, and Cost-Efficient Reinforcement Learning over Decentralized Network}

\author{
    \rm{
        Chaoyi Ruan$^{\text{1}}$\thanks{Chaoyi Ruan and Geng Luo equally contributed to this work} \enskip
        Geng Luo$^{\text{1}}$ \enskip
        Xinyi Wan$^{\text{1}}$ \enskip
        Long Zhao$^{\text{2}}$ \enskip
        Qinghe Wang$^{\text{2}}$
    }
    \\
    \rm{
        Jiaan Zhu$^{\text{3}}$ \enskip
        Duling Xu$^{\text{4}}$ \enskip
        Guanbin Xu$^{\text{5}}$ \enskip
        Dehui Wei$^{\text{1}}$ \enskip
        Xiang Liu$^{\text{1}}$
    }
    \\
    \rm{
        Cheng Li$^{\text{3}}$ \enskip
        Haifeng Sun$^{\text{1}}$ \enskip
        Liang Luo$^{\text{5}}$ \enskip
        Congcong Miao$^{\text{1}}$ \enskip
        Jialin Li$^{\text{1}}$
    }
    \\
    \vspace{0.2in}
    {$^{\text{1}}$\textit{NUS} \enskip
     $^{\text{2}}$\textit{Anhui University} \enskip
     $^{\text{3}}$\textit{USTC} \enskip
     $^{\text{4}}$\textit{Renmin University of China} \enskip
     $^{\text{5}}$\textit{Independent Researcher}}
}

\maketitle

\begin{abstract}

\noindent
LLM reinforcement learning (RL) requires frequent synchronization of large model parameters between the trainer and distributed rollout actors.
High-throughput RL post-training therefore relies on dedicated RDMA HPC/cloud clusters, an infrastructure cost most organizations cannot absorb.
A natural alternative is to aggregate loosely-coupled GPUs over standard Ethernet and WAN links, but this commodity connectivity cannot sustain full-weight broadcasts: synchronizing an 8B model can take over 100~seconds on bandwidth-limited links, while rollout generation typically takes tens of seconds.

Toward making RL practical in this regime, we observe that RL fine-tuning yields highly sparse per-step updates, with only around 1\% of parameter elements changing.
On top of this insight, we present AuroraRL, a novel high-performance RL training system that preserves bit-exact updates without dropping or quantizing information, designed for commodity-networked, loosely-coupled GPU resources. 
AuroraRL represents each step as a sparse delta checkpoint, pipelines delta extraction with multi-stream transmission, overlaps transfer with rollout generation, and coordinates heterogeneous workers with throughput- and bandwidth-aware scheduling plus lease-based fault tolerance.
Across Qwen3 4B--14B models deployed in up to four geographic regions, AuroraRL shrinks per-step weight transfer by 79$\times$ on Qwen3-8B, delivers 1.3--9.5$\times$ higher throughput than dense-broadcast baselines (PrimeRL-Full, async-tolerant, multi-stream variants), and brings end-to-end training within 8.91\% of an ideal RDMA single-datacenter baseline, while transparently tolerating common failures and preserving training accuracy.
By leveraging on-demand, cross-cloud GPUs over commodity links, AuroraRL delivers 1.21--1.59$\times$ higher tokens per dollar than reserved RDMA clusters at comparable throughput.

\end{abstract}

\sloppy

\section{Introduction}
\label{sec:intro}

\begin{figure}[!t]
    \centering
    \includegraphics[width=0.45\textwidth]{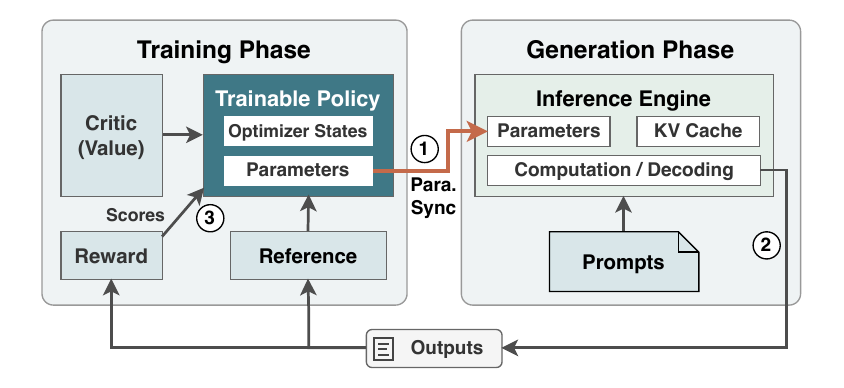}
    \caption{RL training architecture for LLMs. The Trainer holds the policy and auxiliary models; Rollout Actors generate rollouts from prompts. Updated weights are transferred every iteration.}
    \label{fig:rl-arch}
\end{figure}

\noindent
Reinforcement learning (RL) has emerged as a critical post-training paradigm for large language models (LLMs), with systems such as DeepSeek-R1~\cite{guo2025deepseek} and GPT-4o~\cite{openai_gpt4o} demonstrating that RL can unlock strong reasoning capabilities.
RL post-training follows a trainer-actor architecture in which a Trainer updates the policy and refreshes Rollout Actors, who generate samples and return them for training (\autoref{fig:rl-arch}).
In this workflow, the additional transfer of updated parameters to Rollout Actors becomes a new potential bottleneck for end-to-end throughput.
Because RL requires a full-weight refresh at every training step, as the model size grows, this transfer overhead increasingly grows and leaves the compute resources idle while waiting for the next policy version.

State-of-the-art RL systems assume RDMA-connected GPU HPC clusters as the default deployment model, which we refer to as \textit{tightly-coupled clusters}.
In these clusters, RDMA fabrics and co-located scheduling make per-step synchronization across the Trainer and Actors fast and predictable.
The design of existing RL frameworks is fundamentally based on high-bandwidth RDMA fabrics as the backbone.

In contrast, \textit{loosely-coupled deployments} aggregate on-demand GPU resources across multiple cloud providers, university labs, or geographic regions~\cite{lim2024accelerating,skypilot2023,cursor2026composer2} via decentralized networks.
Three factors are pushing academic and frontier industrial users towards this direction. (1) Capacity: The aggregate footprint of a large RL post-training job, including the trainer and geographically distributed inference fleet, can exceed what any single region supplies, as exemplified by Cursor~\cite{cursor2026composer2} which production training spanning three GPU regions and four CPU regions, and by Google Gemini trained across data centers~\cite{gemini2023multi}.
(2) Resilience: Decoupled multi-region deployments absorb regional outages and provider-side disruptions over the multi-day to multi-week timescales of modern RL campaigns.
(3) Accessibility: On-demand commodity GPUs are the most economically viable option for academic groups, startups, and small organizations outside the few hyperscalers.
However, a common constraint in such deployment is network connectivity: cross-region and cross-provider links typically offer only 1--10~Gbps bandwidth and exhibit high WAN latency and loss rate~\cite{lim2024accelerating}.

Existing RL system designs are poor fit for decentralized networks because they couple training progress with frequent dense broadcasts and fine-grained coordination.
Loosely-coupled deployments face two fundamental challenges.
The first is a \textit{decentralized network barrier}, where limited bandwidth and high jitter stall both data movement and control loops.
The second challenge is \textit{heterogeneity and instability} from variable node throughput and preemption.
Prior RL systems such as OpenRLHF~\cite{hu2024openrlhf}, veRL~\cite{sheng2025hybridflow}, and StreamRL~\cite{zhong2025streamrl} optimize rollout efficiency, but rely on high-bandwidth networks, e.g., RDMA fabric or dedicated cross-DC $\geq$80~Gbps links.
When these RL systems are ported to a decentralized network, the full-weight broadcast suffers from low effective throughput. This causes the synchronization of a small 8B model to take over 100~seconds (\autoref{tab:sync-cost-example}), far exceeding the time required for rollout generation and severely degrading GPU utilization.

Optimizing distributed RL on decentralized networks requires a fundamental shift in design principles.
Rather than treating the RL training loop as a \emph{black box} and relying solely on high-bandwidth hardware to brute-force expensive synchronization, we argue for a \textit{white box} approach that reveals the structures in RL workflows and co-designs system-level mechanisms with algorithmic insights.

Through extensive profiling across multiple model families and RL algorithms, we discover a critical property that makes bandwidth-efficient RL training feasible: \textit{parameter updates are remarkably sparse}.
Unlike pre-training, only 1--3\% of parameter elements change per iteration in existing models, creating an opportunity to reduce transfer volumes by two orders of magnitude.
This inherent sparsity fundamentally changes the bandwidth demand for distributed RL.

Inspired by this principle, we build \sys, a novel RL post-training system designed for decentralized GPU pools, delivering high throughput and cost-efficiency without sacrificing accuracy.
\sys follows the common practice of one-step asynchronous lag in recent popular RL systems~\cite{zhong2025streamrl,noukhovitch2025async,fu2025areal}, with three new co-designed mechanisms underpinning the workflow.
First, \sys unifies checkpoint storage and network transfer into a single abstraction, the \textit{lossless delta checkpoint}.
Each training step produces a versioned, immutable artifact that captures only the sparse parameter changes with full precision, using delta-encoded variable-length indexing to further compress position metadata beyond the base sparsity ratio.
Second, a \textit{streaming transfer protocol} pipelines delta extraction with cut-through forwarding over multiple parallel network streams.
The protocol also employs relay-based fanout to scale delivery across regions, while overlapping transfer with rollout generation so that Actors stage the next version behind ongoing computation.
Third, \textit{heterogeneity-aware scheduling} combined with \textit{lease-based fault tolerance} coordinates loosely-coupled workers.
The scheduler sizes each Actor's batch in proportion to its observed throughput and link bandwidth so that all Actors complete within a one-step policy lag.
It applies time-bounded leases to detect failures implicitly, and redistribute work without global barriers while preventing stale rollouts.

We implement \sys atop FSDP2~\cite{pytorch-fsdp2}  and vLLM~\cite{kwon2023efficient} as high-throughput training and inference engines, requiring no modifications to the underlying RL algorithms.
We evaluate \sys on Qwen3 models from 4B to 14B across up to four geographic regions spanning North America, Europe, and Asia-Pacific.
\sys reduces per-step transfer payload by 79$\times$ for Qwen3-8B and improves throughput by 2.4--9.5$\times$ over full-weight broadcast across WAN (1.3--6.1$\times$ over PrimeRL's async and multi-stream variants), narrowing the gap to an ideal RDMA single-datacenter baseline to within 8.91\%.
Leveraging cross-cloud GPUs over commodity links, \sys achieves 1.21--1.59$\times$ higher tokens per dollar than RDMA clusters while transparently tolerating common failures, like GPU faults, spot preemptions, etc.

\section{Background and Motivation}
\label{sec:back}

\begin{figure*}[!t]
	\centering
	\includegraphics[width=0.8\textwidth]{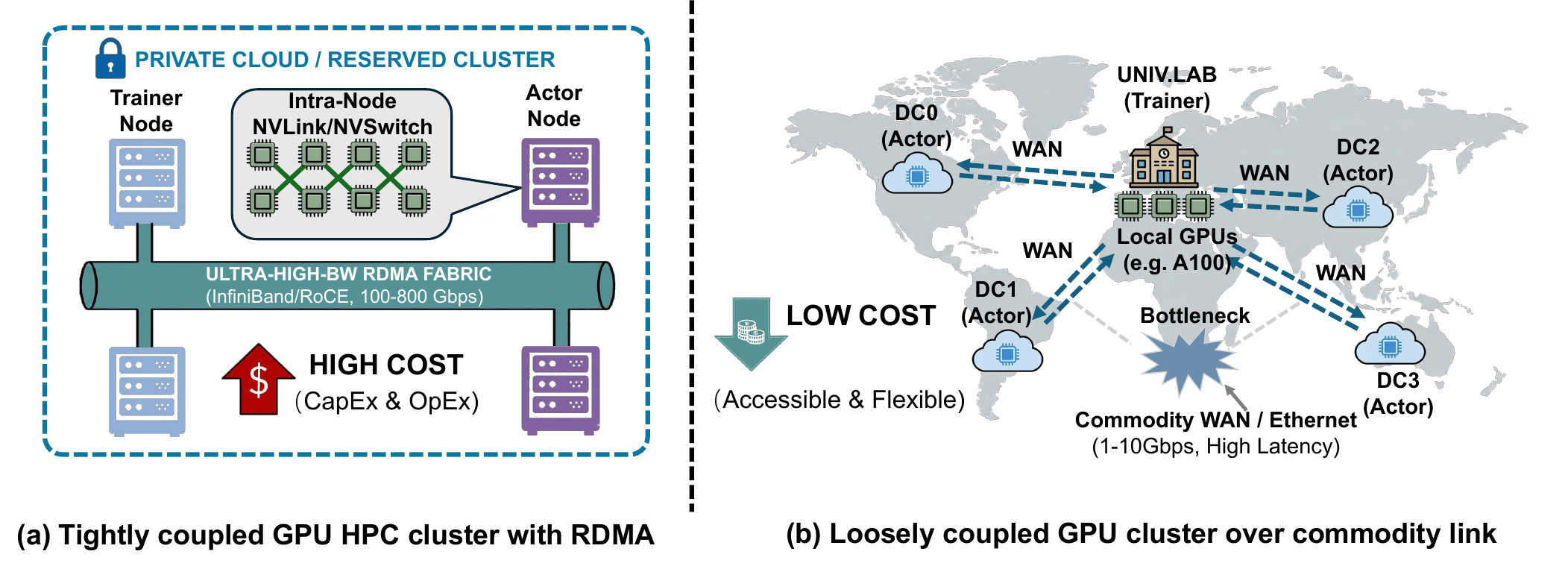}
	\caption{Two paradigms for RL training. Left: tightly coupled clusters with RDMA interconnects (100--800~Gbps) and high cost. Right: loosely coupled resources across labs and clouds, connected via cross-cloud links (1--10~Gbps).}
	\label{fig:paradigm}
\end{figure*}

\subsection{Reinforcement Learning for LLMs}
\label{sec:rl-arch}

\noindent
Reinforcement learning has emerged as the standard paradigm for post-training LLMs, enabling capabilities beyond supervised fine-tuning~\cite{nakano2021webgpt, stiennon2020learning}.
Most systems follow a training-generation workflow (\autoref{fig:rl-arch}) that tightly couples a Trainer with a set of Rollout Actors.
Each step comprises three actions: \ding{172} the Trainer synchronizes the updated policy parameters $\pi_t$ to the Rollout Actors, \ding{173} the workers generate rollout samples from prompts and return trajectories, and \ding{174} the Trainer computes rewards and advantages using the reward model and critic, then updates the policy to produce $\pi_{t+1}$.
In this workflow, parameter transfer is on the critical path of each step.
To mitigate this cost, recent systems adopt one-step asynchronous RL~\cite{zhong2025streamrl,noukhovitch2025async,fu2025areal}, where the generation at step $t{+}1$ proceeds with stale weights $\pi_t$ while the Trainer computes $\pi_{t+1}$ in parallel.
This shifts synchronization off the critical path by overlapping it with generation, provided that the transfer completes within one generation period.

Although rollout samples returned from generation to training are very small, weight transfer incurs a high communication cost.
Even an 8B model in BF16 requires sending approximately 16~GB to \textit{each} Rollout Actor per step, and this cost scales linearly with model size.

\subsection{The HPC Cloud Premium}
\label{sec:gap}

\noindent
State-of-the-art RL systems assume an RDMA-connected HPC cluster within a single datacenter as the default deployment model, as illustrated in \autoref{fig:paradigm}-(a).
With high-end RDMA fabrics and co-located scheduling, a full-weight refresh completes in roughly one second (\autoref{tab:sync-cost-example}).
For example, Perplexity~\cite{perplexity2025weight-transfer} reports similar sub-2~s transfers in RL post-training deployments, so existing designs treat per-step synchronization as a small overhead.

This single-datacenter HPC model introduces both cost and access barriers (\autoref{tab:gpu-cost}), and---as we show next---a hard capacity ceiling that the largest RL jobs already exceed.
RDMA clusters typically charge a network premium over commodity Ethernet, for example \$39/hour versus \$31/hour for 16$\times$H100 on Hyperbolic, which is a 26\% increase~\cite{hyperbolic2025}.
Providers also gate these clusters behind minimum reservations, so an exploratory 48-hour run can become a one-week commitment that inflates cost by 3.5$\times$.
Beyond cost, access remains constrained. Academic institutions, government labs, and small organizations often lack the procurement relationships or credit lines needed to reserve large GPU blocks.

\begin{table}[t]
\centering
\caption{Cost and accessibility trade-offs for renting a 16-GPU deployment: RDMA clusters typically require long minimum commitments and sales-team approval at scale, whereas cross-cloud configurations can stitch together on-demand, flexible capacity across providers.}
\label{tab:gpu-cost}
\resizebox{0.48\textwidth}{!}{%
\begin{tabular}{@{}lccc@{}}
\toprule
\textbf{Configuration (Provider)} & \textbf{\$/hr} & \textbf{Min.} & \textbf{BW} \\
\midrule
\multicolumn{4}{@{}l}{\textit{Tightly coupled (RDMA fabric, reserved)}} \\
\;\;2$\times$8$\times$H100, EFA (AWS)        & 32 & 24\,hr & 3.2\,Tbps \\
\;\;2$\times$8$\times$H100, IB (Lambda)      & 38 & 1\,wk  & 3.2\,Tbps \\
\midrule
\multicolumn{4}{@{}l}{\textit{Loosely coupled (cross-cloud, on-demand)}} \\
\;\;8$\times$H100 (Hyperbolic\,\cite{hyperbolic2025}) + 8$\times$H100 (Hyperstack\,\cite{hyperstack2025}) & 27 & 1\,hr & 1\,Gbps \\
\;\;8$\times$H100 (Hyperbolic\,\cite{hyperbolic2025}) + 8$\times$A100 (Lambda\,\cite{lambda-cluster})    & 26 & 1\,hr & 1\,Gbps \\
\midrule
\multicolumn{4}{@{}l}{\textit{Loosely coupled (spot / preemptible)}} \\
\;\;2$\times$8$\times$H100 spot (GCP a3\,\cite{gcp-cloud-pricing-calculator})\textsuperscript{$\dagger$} & $\sim$64 & 1\,hr & 3.2\,Tbps \\
\bottomrule
\end{tabular}}

{\scriptsize\raggedright \textsuperscript{$\dagger$}\,Spot rate is $\sim$50\% off the on-demand price of \$176/hr; subject to preemption.\par}
\end{table}

\begin{table}[t]
\centering
\caption{Impact of network bandwidth on full-model synchronization for Qwen3-8B (16~GB in BF16).}
\label{tab:sync-cost-example}
\resizebox{0.48\textwidth}{!}{%
\begin{tabular}{lcccc}
\toprule
\textbf{Network} & \textbf{Trainer} & \textbf{Actor} & \textbf{BW} & \textbf{Sync} \\
\midrule
HPC fabric (RDMA) & \multirow{2}{*}{40~s} & \multirow{2}{*}{45~s} & 100~Gbps & 1.3~s  \\
Commodity network    &                     &                      & 1~Gbps   & 128~s \\
\bottomrule
\end{tabular}
}
\end{table}

\subsection{Opportunity: Decentralized GPU Pools}
\label{sec:opportunities}

\para{Decentralized GPU pools: capacity, resilience, and accessibility.}
Three converging forces---capacity ceilings, multi-region resilience, and broad accessibility---push academic groups, startups, and frontier industry labs alike toward decentralized GPU pools.

Cursor's Composer~2, a 1.04T-parameter agentic-RL training job, spans three GPU regions and four CPU regions~\cite{cursor2026composer2} because no single region can host its trainer, distributed inference fleet, and hundreds of thousands of stateful environment pods. The decoupled topology also let the job survive partial outages of inference and environment services without restarting, making decentralized RL a structural requirement at scale.

Three deployment patterns recur in practice and serve users along all three axes: (1) heterogeneous instances within a single cloud, where training and generation run on different GPU types over commodity networking~\cite{mlaas2022, jakiro2025, zhong2025streamrl, MLSYS2022_0cafb789}; (2) cross-region or multi-cloud RL that aggregates on-demand GPUs from providers such as AWS, Lambda, and GCP~\cite{erben2024traindeeplearningmodels,he2025hetrlefficientreinforcementlearning,cursor2026composer2,aws-ec2-price-calculator,lambda-cluster,gcp-cloud-pricing-calculator}; and (3) cross-institution collaborations that pool GPUs across labs~\cite{ryabinin2020crowdsourcedtraininglargeneural, borzunov2023petalscollaborativeinferencefinetuning}, all illustrated in \autoref{fig:paradigm}-(b). As \autoref{tab:gpu-cost} shows, combining 16 on-demand GPUs across providers costs \$26 to \$27 per hour with 1-hour billing versus \$32 to \$38 per hour for an RDMA cluster with 24-hour to one-week reservations, but commodity links deliver only 1 to 10~Gbps rather than multi-terabit RDMA fabrics.
A fourth option is preemptible spot capacity at major hyperscalers, which offers up to $\sim$60\% off on-demand for the same high-bandwidth infrastructure (e.g., GCP H100 spot~\cite{gcp-cloud-pricing-calculator}), at the cost of frequent reclamation: a spot instance typically survives only a few hours before being taken back~\cite{wu2025rlboost}. A multi-day RL run therefore brings many such events per Actor, so the system must absorb failures continuously rather than treat them as rare exceptions.

Recent decentralized-GPU work focuses on LLM training, reducing transfer bandwidth by exploring natural or induced top-$k$/threshold sparsification, with error feedback or bounded staleness used to contain drift on the induced path. These techniques apply either to checkpoint-to-storage traffic~\cite{eisenman2022checknrun, yao2025lowdiff} or to symmetric gradient synchronization on datacenter~\cite{fei2021omnireduce}, RDMA~\cite{wang2025zen}, and WAN~\cite{lim2024accelerating} fabrics.

\subsection{Challenges and Existing RL Systems}
\label{sec:existing-systems}

\noindent
However, none of the above targets RL post-training. In contrast, RL post-training over decentralized GPUs broadcasts each new policy from Trainer to Actors on the generation critical path. This asymmetric pattern surfaces three challenges that no existing RL architecture handles jointly.

\para{Challenge 1 (C1): the commodity network barrier.}
Decentralized deployments connect the Trainer and Actors over commodity networks (1--10~Gbps cross-region, cross-cloud, or heterogeneous-instance links), where full model weight transfer dominates iteration time and throttles throughput regardless of GPU speed.
At datacenter-grade bandwidths (100+~Gbps), weight transfer is a small fraction of each iteration. However, on commodity links, it stretches the critical path.
As \autoref{tab:sync-cost-example} shows, a 16~GB model transfers in 1.3~s on an HPC fabric, within a 45~s generation window. On a 1~Gbps link, the same transfer takes 128~s, beyond what even one-step asynchronous RL can hide.

\para{Challenge 2 (C2): hardware and network heterogeneity.}
Cross-cloud deployments aggregate nodes with substantial variation in compute capacity and network bandwidth.
GPU instances range from older L40 and A100 GPUs to newer H100 hardware, with generation throughput differing by 2 to 3$\times$.
Network conditions are equally heterogeneous: Actors from nearby providers may achieve 5 to 10~Gbps bandwidth to the Trainer, while those across continents operate at 1 to 3~Gbps.
Stragglers from thermal throttling or resource contention further shift these rates during a run.

\para{Challenge 3 (C3): dynamic membership and failures.}
Loosely coupled deployments must treat membership change as routine rather than exceptional.
Actors disappear involuntarily through spot preemption (multi-percent per hour for H100), hardware or network failures, and regional outages; and join or leave deliberately as operators add capacity, shift regions, or resize the pool mid-run.
Each event can occur at any moment within a multi-day to multi-week RL campaign, and the system must absorb them without pausing ongoing rollout generation or invalidating in-flight work.

\begin{table}[t]
\centering
\caption{Comparison of RL system capabilities for geo-distributed deployment. $\circ$: partial support with significant trade-offs.}
\label{tab:system-comparison}
\small
\begin{tabular}{@{}p{0.22\linewidth}p{0.42\linewidth}ccc@{}}
\toprule
\textbf{Architecture} & \textbf{Example system(s)} & \textbf{C1} & \textbf{C2} & \textbf{C3} \\
\midrule
\textbf{Colocated} & OpenRLHF~\cite{hu2024openrlhf}, veRL~\cite{sheng2025hybridflow} & \ding{55} & \ding{55} & \ding{55} \\
\textbf{Disaggregated} & StreamRL~\cite{zhong2025streamrl} & \ding{55} & $\circ$ & \ding{55} \\
\textbf{Decentralized} & PrimeRL~\cite{prime2025intellect} & \ding{55} & \ding{51} & $\circ$ \\
\midrule
\textbf{Sparse-delta} & \textbf{\sys} & \ding{51} & \ding{51} & \ding{51} \\
\bottomrule
\end{tabular}
\end{table}

\para{Why do existing systems fall short?}
As \autoref{tab:system-comparison} shows, no prior architecture handles all three challenges. \textbf{Colocated} systems~\cite{hu2024openrlhf,sheng2025hybridflow} assume datacenter fabrics, homogeneous hardware, and stable membership; \textbf{disaggregated} systems~\cite{zhong2025streamrl} tolerate some heterogeneity but require $\sim$80~Gbps dedicated links and a stable cluster; \textbf{decentralized} systems~\cite{prime2025intellect} reach internet scale and absorb churn but still broadcast full weights, hiding that cost via multi-step staleness that compromises model quality.

\section{Key Insight and Design Principles}
\label{sec:challenges}

\begin{figure}[!t]
\centering
\includegraphics[width=0.45\textwidth]{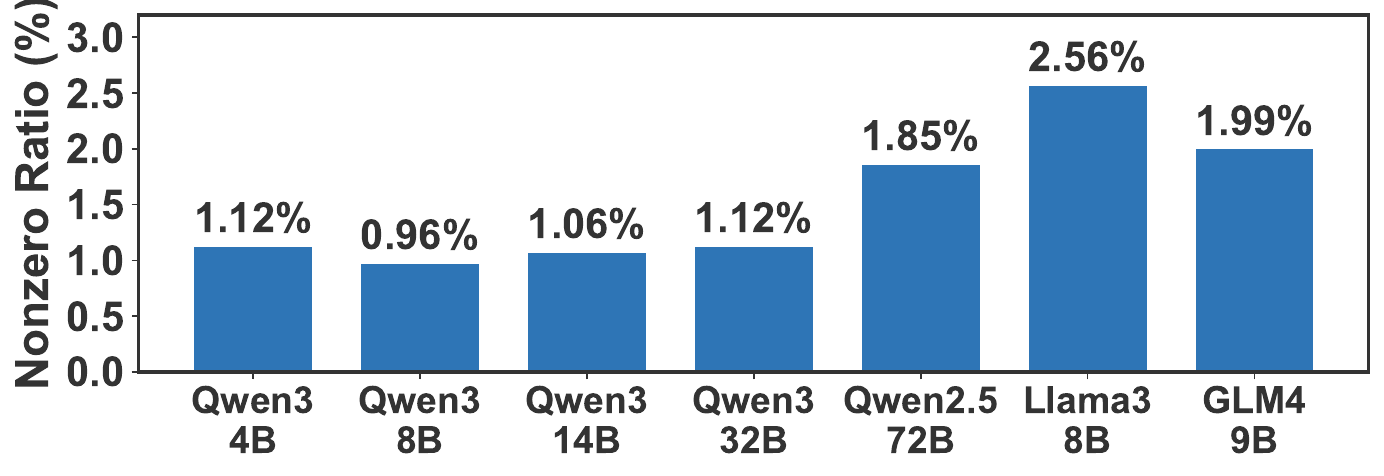}
\caption{Fraction of nonzero parameter updates after one RL step across different models.}
\label{fig:nonzero-fraction}
\end{figure}

\begin{table}[!t]
\centering
\small
\caption{Fraction of nonzero parameter updates after one RL step for Qwen3-8B under different algorithms.}
\label{tab:nonzero-algo}
\begin{tabular}{lccc}
\toprule
\textbf{Algorithm} & \textbf{GRPO}~\cite{shao2024deepseekmath} & \textbf{RLOO}~\cite{ahmadian2024basicsrevisitingreinforcestyle} & \textbf{OPO}~\cite{hao2025onpolicyrloptimalreward} \\
\midrule
\textbf{Nonzero Ratio $\rho$ (\%)} & 0.96 & 0.93 & 1.06 \\
\bottomrule
\end{tabular}
\end{table}

\begin{figure*}[t]
    \centering
    \includegraphics[width=\linewidth]{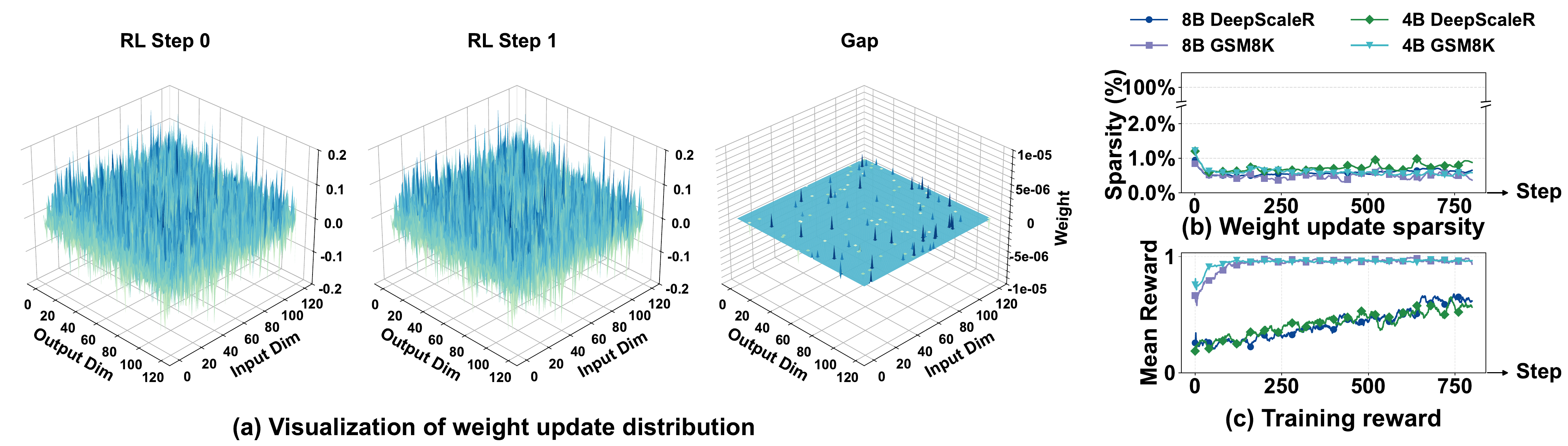}
    \caption{Analysis of training dynamics.
    \textbf{(a)} Visualization of weight update distribution, demonstrating the sparse nature of parameter updates.
    \textbf{(b)} Weight update sparsity and \textbf{(c)} training reward throughout RL training. We train 4B/8B models on GSM8K~\cite{cobbe2021gsm8k} and DeepScaleR~\cite{deepscaler2025} datasets for 800 rollout steps with a fixed learning rate of $1\times10^{-6}$.}
    \label{fig:combined-analysis}
\end{figure*}

\noindent
The challenges in \autoref{sec:existing-systems} appear fundamental if we treat the RL system as a monolithic black box.
However, by dissecting the RL workflow, we uncover a critical property: \textit{parameter update sparsity}.
This property, combined with the structural independence of RL tasks, allows us to transform network bottlenecks into design opportunities.
We propose three principles to address the identified challenges.

\para{Empirical basis: fine-grained sparsity.}
We first quantify parameter changes in a single RL step. An LLM consists of $K$ parameter tensors $\{W^{(k)}\}_{k=1}^{K}$ (e.g., attention projections and MLP weights). Let $W^{(k)}_{t}$ and $W^{(k)}_{t+1}$ denote the values of tensor $k$ before and after one RL step, and define the element-wise difference $\Delta W^{(k)} \!=\! W^{(k)}_{t+1} - W^{(k)}_{t}$ obtained by differencing consecutive checkpoints. We define the element-wise \textit{nonzero ratio} $\rho$ across the entire model as
\begin{equation}
\rho = \frac{1}{\sum_{k=1}^{K} |W^{(k)}|}\sum_{k=1}^{K} \|\Delta W^{(k)}\|_0
\label{eq:nonzero-ratio}
\end{equation}
where $|W^{(k)}|$ is the number of scalar parameters in tensor $W^{(k)}$ and $\|\cdot\|_0$ counts nonzero elements. Crucially, this sparsity is fine-grained: while almost every tensor in the model receives updates, only a tiny subset of elements within each tensor changes, as shown in \autoref{fig:combined-analysis}-(a).
This ratio is consistently low across different model families and under different RL algorithms. For instance, \autoref{fig:nonzero-fraction} shows that for Qwen3-4B, only 1.12\% of parameters change after an update.
Similar nonzero ratios are observed for larger models such as Llama3-8B (2.56\%), GLM4-9B (1.99\%), and Qwen2.5-72B (1.85\%).
In addition, \autoref{tab:nonzero-algo} shows that, when measured on Qwen3-8B, RL algorithms like GRPO~\cite{shao2024deepseekmath}, RLOO~\cite{ahmadian2024basicsrevisitingreinforcestyle}, and OPO~\cite{hao2025onpolicyrloptimalreward} all modify around 1\% of parameters per step.

Importantly, this sparsity is uniformly distributed across all weight tensors: every layer receives updates, but only a small fraction of elements within each tensor are modified.
This pattern is not coincidental but arises from fundamental properties of RL fine-tuning. The learning rate for post-training alignment is on the order of $10^{-6}$~\cite{jin2025search, yang2024qwen2, liu2025understanding}, two orders of magnitude below the $10^{-4}$ rates used in pre-training~\cite{dubey2024llama, liu2024deepseek}, enabling behavioral refinement while preserving foundational knowledge. Recent theoretical work attributes this behavior to RL's in-distribution nature: online updates are more robust to forgetting and naturally induce smaller parameter changes than supervised approaches~\cite{shenfeld2025rl, mukherjee2025reinforcement}. Regularization techniques including KL divergence constraints and gradient clipping~\cite{zhanggradient} further bound update magnitudes. Crucially, this sparsity is a persistent property rather than a transient artifact: as shown in \autoref{fig:combined-analysis}-(b-c), the update ratio remains stable throughout training, rapidly falling below 1\% and staying there across 800 steps.

Quantization effects further contribute to the observed sparsity. In practical RL pipelines, rollout models are executed in reduced precision such as BF16, as adopted by major RL frameworks including VeRL~\cite{sheng2025hybridflow}, PrimeRL~\cite{prime2025intellect}, and SlimeRL~\cite{slime_github}.
Concretely, training may run the optimizer in FP32 on master weights, but the checkpoints consumed by Actors are represented in BF16. Consequently, small FP32 updates below the local BF16 resolution are not visible to Actors, so \sys's deltas and reported sparsity are defined over these BF16 checkpoints.
Although FP16 provides higher mantissa precision than BF16, it has a substantially smaller dynamic range and a larger minimum representable value above zero \cite{qi2025defeating}, and is less common in RL pipelines because its limited range increases gradient-underflow risk and requires loss scaling \cite{micikevicius2017mixed}. The detailed numerical analysis is in \autoref{sec:appendix}.

The above empirical study inspires us to rethink the RL workflow from the perspective of sparsity (\autoref{sec:overview} and \autoref{sec:design}).

\section{Overview of \sys}
\label{sec:overview}

\begin{figure}[t]
    \centering
    \includegraphics[width=0.47\textwidth]{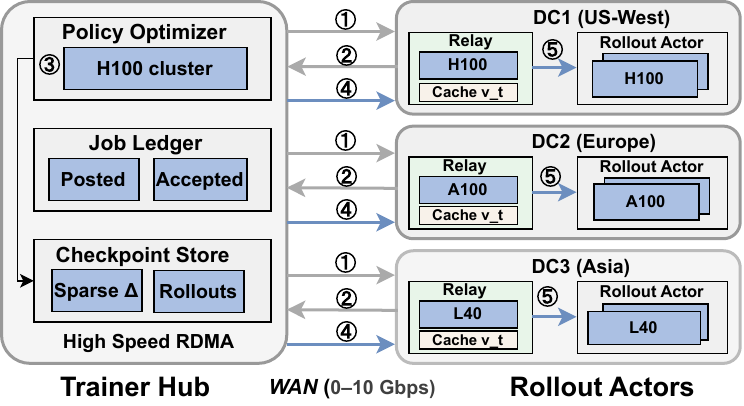}
    \caption{\sys architecture overview. The Trainer Hub posts jobs and sparse deltas; Rollout Actors claim prompts, generate samples, and return results. Each region contains a Relay that receives deltas from the Trainer and forwards them to peer Rollout Actors.}
    \label{fig:system-arch}
\end{figure}

\noindent
We propose \sys, an RL training system for LLMs that addresses three commodity-network barriers (\autoref{sec:existing-systems}) by exploiting three properties of RL workflows, without sacrificing model accuracy: \textit{fine-grained sparsity} of per-step weight updates ($\sim$1\%, \autoref{sec:challenges}) compresses weight transfers (C1); \textit{decomposability} of sparse deltas enables streaming overlap with rollout generation that hides WAN latency (C1); and \textit{independence} of rollout tasks enables elastic per-task scheduling and publish--subscribe delta dissemination, absorbing heterogeneity and dynamic membership (C2, C3).

\para{Deployment model.}
\sys targets a deployment where a centralized Trainer runs on a well-provisioned cluster while inference capacity is distributed across multiple cloud providers or geographic regions. \autoref{fig:system-arch} illustrates the resulting architecture. The Trainer Hub connects to regional Relays over cross-cloud links up to 10~Gbps, depending on provider peering and network conditions.

\para{System components.}
The architecture comprises two logical tiers (\autoref{fig:system-arch}). The \textbf{Trainer Hub} serves as the coordinator for policy state. It runs the policy optimizer (e.g., GRPO) on a well-provisioned cluster (e.g., H100s) interconnected via high-speed RDMA fabric, maintains a \textbf{Job Ledger} tracking posted and accepted work, and operates a \textbf{Checkpoint Store} holding versioned sparse deltas and collected rollouts.
\textbf{Rollout Actors} span heterogeneous hardware across geographic regions (H100, A100, and L40 clusters in the figure); they receive policy deltas, claim prompts, generate rollouts locally, and return results. In each region, one Actor is designated as the \textbf{Relay}, a dual-role node that both generates rollouts and serves as a regional proxy. The Relay caches the current policy version $v_t$ and forwards deltas to peer Actors in its region, reducing cross-region traffic from $O(N)$ transfers to one per region.

\para{Workflow.}
\sys enforces a strict one-step policy lag~\cite{zhong2025streamrl}, bounding staleness while allowing rollouts, update transfer, and version preparation to overlap across regions.
In \autoref{fig:system-arch}, the data plane carries two types of traffic: gray arrows denote rollout traffic and blue arrows denote sparse delta transfer.
The control plane handles job coordination through time-bounded leases: the Job Ledger issues prompts to Actors, and Actors later submit results.
Each iteration proceeds through five stages, as annotated in \autoref{fig:system-arch}. The Job Ledger issues prompts to Actors  (\ding{192}), which generate rollouts on their active version $\pi_v$. Completed rollouts flow back (\ding{193}) to the Checkpoint Store, where they are aggregated for training. The optimizer consumes the collected rollouts and produces the updated policy $\pi_{v+1}$ (\ding{194}). The update is encoded as a sparse delta $\mathcal{D}_{v+1}$, stored in the Checkpoint Store (\ding{195}), and transferred outward through Relays. Relays forward $\mathcal{D}_{v+1}$ (\ding{196}) and Actors activate the new version between batches, while laggards catch up asynchronously without blocking other regions.
Lease expiry automatically returns orphaned prompts to the pool, avoiding global stalls on stragglers.

\section{\sys's System Design}
\label{sec:design}

\noindent
This section describes four mechanisms within \sys: sparse delta checkpoints (\autoref{sec:delta-checkpoints}), streaming delta transfer (\autoref{sec:overlapped-transfer}), heterogeneity-aware scheduling (\autoref{sec:scheduling}), and fault tolerance with elastic scaling (\autoref{sec:fault}).

\subsection{Lossless Sparse Delta Geo-Checkpoints}
\label{sec:delta-checkpoints}

\noindent
Distributed training frameworks typically decouple parameter persistence from synchronization, saving dense model checkpoints to storage while broadcasting updates over RDMA or NVLink protocols. This separation creates consistency challenges in geo-distributed deployments where network faults are common. If a transfer is interrupted or retried, the system struggles to verify whether a remote Actor has successfully applied the update. As a result, matching each rollout back to its policy version becomes a stateful tracking problem rather than a direct check.

\sys addresses this by unifying checkpoint storage and network transfer into a single abstraction: the \textit{delta checkpoint}. Rather than sending ephemeral updates, the Trainer produces a versioned, immutable file $\mathcal{D}_v$ for each step, complete with a unique identifier and integrity hash. This design embeds the version control inherent to storage systems directly into the transport layer. By treating network transfer as the replication of a persistent artifact, \sys ensures that partial failures never result in ambiguous states. This unification simplifies the distributed protocol, naturally supporting safe Relay caching, peer-assisted fanout, and verifiable acceptance predicates (\autoref{sec:fault}).

\begin{figure}[t]
\centering
\includegraphics[width=0.46\textwidth]{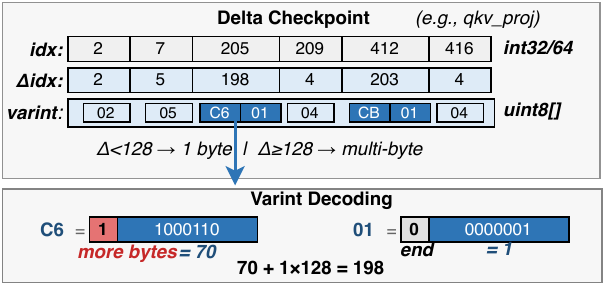}
\caption{Delta checkpoint index encoding.
First, absolute indices (\texttt{int32/64}) are replaced with delta offsets ($\Delta$idx) relative to the previous index.
Then, each offset is encoded as a variable-length unsigned byte sequence: offsets $<128$ take one byte, while larger offsets use multiple bytes with a high-bit continuation flag.}
\label{fig:delta_representation}
\end{figure}

\para{Sparse encoding.}
As illustrated in \autoref{fig:delta_representation}, the Trainer flattens each tensor's delta into a one-dimensional index space and stores the non-zeros as two 1D arrays, \texttt{idx} and \texttt{val}.
For attention and MLP projections, \sys writes deltas under fused inference names by stacking split HuggingFace blocks in a fixed order. For example, Q, K, and V updates become a single \texttt{qkv\_proj} delta by adding deterministic block offsets to each component's linear indices. Similarly, Gate and Up matrices fuse into \texttt{gate\_up\_proj}.
Actors apply the update with a flat scatter-add over the parameter's storage, keeping the payload directly consumable by the inference engine.

A naive encoding represents each nonzero update as a fixed-width (index, value) pair, with an \texttt{int32} or \texttt{int64} index (depending on tensor size) and a \texttt{bfloat16} value, thereby dedicating two-thirds or more of the payload to position metadata.
\sys reduces this overhead with two complementary techniques.
First, it applies \textit{delta encoding} to the sorted index array: the encoder stores the first index as-is, then replaces each subsequent index with its difference from the predecessor. Since non-zero updates are scattered across the tensor, most index differences are small and would fit within an 8-bit unsigned integer. However, rare large gaps make a fixed-width \texttt{uint8} insufficient.
To efficiently capture this long-tailed distribution without over-provisioning every entry, \sys encodes the delta sequence using unsigned LEB128, a variable-length integer representation in which the most significant bit of each byte indicates continuation.
Differences smaller than 128 therefore occupy a single byte, while larger gaps extend to multiple bytes. For example, the value 198 is encoded as two bytes (\texttt{C6 01}). The first byte \texttt{C6} (\texttt{1100 0110}\textsubscript{2}) carries payload 70 and sets the continuation bit, indicating that additional bytes follow. The second byte \texttt{01} (\texttt{0000 0001}\textsubscript{2}) carries payload 1 as the final byte. Interpreting the payload in LEB128 yields $70 + (1 \ll 7) = 198$ (\autoref{fig:delta_representation}).
Together, these techniques reduce the index footprint from four bytes per entry to fewer than two on average, further cutting total checkpoint size by 30--50\%.

\para{Lossless precision.}
Unlike gradient compression techniques~\cite{lim2024accelerating} that trade accuracy for bandwidth through lossy quantization, \sys employs lossless compression.
The non-zero parameter based delta-encoded varint representation preserves full precision,
ensuring Actors reconstruct exactly the same checkpoint as under full-checkpoint transmission.

\subsection{Streaming Delta Transfer Protocol}
\label{sec:overlapped-transfer}

\noindent
Sparse encoding reduces transfer volume, but transfer must still complete quickly enough to keep Actors within \sys's one-step lag bound. This creates a per-iteration deadline: each Actor should finish staging $\mathcal{D}_{v+1}$ before it completes its current rollout batch on $\pi_v$, otherwise it will either idle waiting for the next version or fall behind and become unable to contribute valid rollouts.

In our US--Canada deployment with Qwen3-8B in \autoref{sec:eval}, extracting a sparse delta takes approximately 5~s, while transferring a 202~MB delta over a single TCP stream takes 4.71~s. Even after sparsity reduces payload size, the WAN transport is still sensitive to large bandwidth-delay products and loss, and a single stream commonly leaves capacity unused due to congestion control and head-of-line blocking. \sys therefore treats each delta checkpoint as a stream and accelerates transfer with two techniques targeting tail latency. First, it pipelines extraction with cut-through forwarding, allowing transmission to begin before the full delta is materialized.
Second, it stripes segments across multiple parallel TCP streams so that loss-induced stalls on one stream do not block the others, keeping the link busy under transient loss.

\begin{figure}[t]
\centering
\includegraphics[width=0.48\textwidth]{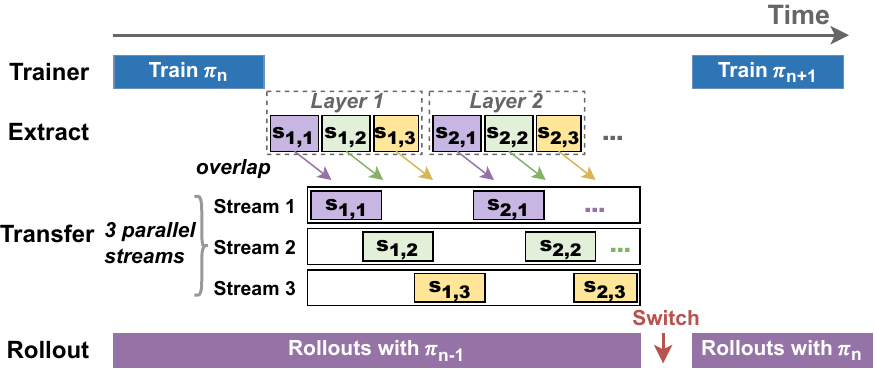}
\caption{Pipelined delta extraction and multi-stream transfer ($S{=}3$). Segments are striped round-robin across parallel streams, overlapping extraction with transfer. Actors continue rollouts on $\pi_{n-1}$ and switch to $\pi_n$ when transfer completes.}
\label{fig:prepare-activate}
\end{figure}

\para{Overlapped extraction and transfer.}
As illustrated in \autoref{fig:prepare-activate}, the Trainer extracts the sparse delta by scanning parameters and encoding nonzeros. Instead of treating $\mathcal{D}_v$ as a monolithic file, \sys packetizes it into a sequence of segments that can be transmitted and buffered independently and reassembled deterministically, with integrity verified against the delta checkpoint hash. The Trainer emits each segment immediately after encoding it, and Relays forward segments on arrival, creating cut-through overlap between extraction, cross-region transfer, and intra-region fanout.

\sys uses $S$ parallel TCP streams and stripes segments round-robin across them. Segment-level striping serves two purposes. First, it improves link utilization on WAN paths where a single TCP stream underutilizes bandwidth due to conservative congestion control. Second, it reduces long-tail sensitivity to loss: a retransmission stall on one stream delays only its assigned segments, while other streams continue to make progress. This granularity also avoids imbalance under skewed sparsity patterns where a small subset of layers carries most of the delta bytes. In our deployment, multi-stream reduces per-step transfer time from 4.71~s to 2.90~s without changing payload size. Finally, \sys overlaps transfer with rollout generation: Actors stage $\mathcal{D}_{v+1}$ while generating on $\pi_v$, so activation can occur at the next safe point instead of blocking on transfer.

\para{Relay-based fanout.}
Multi-stream transfer accelerates delivery to a single receiver, but transfer must also scale to many Actors per region without imposing $O(N)$ cross-region transfers. \sys therefore adopts a relay-based, two-tier push transfer path. For each remote region, the Trainer streams $\mathcal{D}_v$ only to a designated seed Actor, which acts as a relay and proactively forwards the staged update to the peer Actors in that region. The relay forwards blocks as they arrive, overlapping cross-region reception with intra-region fanout.
\sys separates \textit{staging} from \textit{activation}: once the delta is fully staged, the Job Ledger sends a commit command for version $v$ to the relay, and the relay propagates the same commit to its peers so that all Actors switch to $v$ at an end-of-batch safe point via in-place sparse scatter-add, ensuring rollouts never observe a partially applied policy. To preserve consistency under retries and failures, each delta is tagged with a base version, and an Actor accepts/activates $\mathcal{D}_v$ only if its currently active version matches the declared base, preventing out-of-order application.

\subsection{Heterogeneity-Aware Scheduling}
\label{sec:scheduling}

\noindent
Actors vary widely in compute capacity (A100 vs.\ H100 differs by 2--3$\times$) and network bandwidth, and these gaps shift over time as links congest, GPUs throttle, and spot instances are preempted. Equal work assignment creates stragglers, and because \sys enforces a one-step policy lag, an Actor that falls two steps behind has its rollouts invalidated.

\sys addresses this with two mechanisms. \textit{Adaptive job allocation} splits each batch proportionally to estimated throughput, implicitly penalizing slow Actors via EMA-updated estimates. \textit{Version-aware scheduling} gates participation on each Actor's version state, ensuring that only Actors capable of running the target policy version receive work. \autoref{alg:scheduling} summarizes the logic.

\begin{algorithm}[t]
\caption{Heterogeneity-Aware Job Scheduling}
\label{alg:scheduling}
\small
\KwIn{Version $v$, batch size $B$, per-Actor throughput $\tau_a$, exclusion decay factor $\alpha$, EMA factor $\beta$}
$T \gets 0$\;\label{line:agg}
\ForEach{registered Actor $a$}{
    \If{$a.\textnormal{ver} = v$ \normalfont\textbf{or} $a.\textnormal{ver} = v{-}1\ $}
    {$T \gets T + \tau_a$
    \label{line:eligible}}
}
\ForEach{registered Actor $a$}{
    \uIf{$a.\textnormal{ver} = v$ \normalfont\textbf{or} $a.\textnormal{ver} = v{-}1\ $}{
        $B_a \gets \lfloor B \cdot \tau_a / T \rfloor$\;\label{line:full-batch}
        \If{$a.\textnormal{ver} = v{-}1$\label{line:activate}}{
            send \textnormal{\textsc{Commit}}$(v)$ to $a$\;\label{line:commit}
        }
        dispatch $B_a$ requests to $a$\;\label{line:dispatch}
    }
    \lElse{$B_a \gets 0,\; \tau_a \gets \alpha\,\tau_a$\label{line:exclude}}
}
\textbf{On settle:} $\tau_a \gets \beta\,\tau_a + (1{-}\beta)\!\cdot\!(\text{tokens}/\text{elapsed})$\;\label{line:ema}
\end{algorithm}

\para{Adaptive job allocation.}
The user specifies a total batch size $B$ per training step, and the Trainer splits $B$ across eligible Actors in proportion to their estimated generation throughput so that all Actors finish at approximately the same time. The Trainer maintains a per-Actor throughput estimate $\tau_a$ (tokens/s) and computes the eligible aggregate capacity $T = \sum_{a \in \mathcal{E}} \tau_a$, where $\mathcal{E}$ is the set of Actors whose version state permits participation (\autoref{line:eligible}). Each eligible Actor $a$ then receives $B_a = \lfloor B \cdot \tau_a / T \rfloor$ requests (\autoref{line:full-batch}). For example, an H100 at 5{,}000\,tokens/s and an A100 at 2{,}500\,tokens/s split a batch of 300 requests into 200 and 100, respectively.

Fluctuations in compute and network can happen frequently~\cite{MLSYS2020_eca986d5}.  The scheduler adapts to changing conditions automatically. After each settlement, the Trainer updates $\tau_a$ via exponential moving average (\autoref{line:ema}), blending the historical estimate with the Actor's recent throughput. Because $B_a$ is proportional to $\tau_a / T$, a slower Actor sees its $\tau_a$ drop and consequently its share of the next batch shrink, while faster Actors receive more requests. This single feedback signal captures all sources of slowdown, including GPU throttling, network congestion, and transient resource contention, without requiring separate bandwidth tracking mechanism.

\para{Version-aware scheduling.}
Before assigning a new batch, the Trainer checks each Actor's version state to determine eligibility (\autoref{line:eligible}).
An Actor qualifies if it is already on version $v$, or if it is on $v{-}1$ with delta $v$ being staged.
Actors on $v{-}1$ receive a \textsc{Commit} for version $v$ (\autoref{line:commit}) so they activate the staged delta before generating rollouts.
Actors that are more than one version behind are temporarily excluded (\autoref{line:exclude}) and receive no work for the current step.
This decision avoids partial allocation: the entire batch is distributed only among eligible Actors.
Exclusion applies a decay $\alpha$ to $\tau_a$, so rejoining Actors start conservatively and recover throughput share only after demonstrating sustained performance.

\subsection{Fault Tolerance and Elastic Scaling}
\label{sec:fault}

\noindent
Existing fault-tolerance and elasticity works mostly target distributed training and absorb membership change through local model-state redundancy: pipeline templates~\cite{jang2023oobleck}, neighbor-layer redundancy~\cite{thorpe2023bamboo}, in-memory checkpoints~\cite{wang2023gemini}, proactive reconfiguration~\cite{duan2024parcae}, peer-state pipeline adaptation~\cite{gandhi2024recycle}. These approaches assume tightly coupled peers that can exchange state cheaply, an assumption that breaks down across cross-region links where redundancy itself becomes the bottleneck. They also miss RL's actor-learner coupling, in which the Trainer--Actor link publishes versioned policy artifacts rather than synchronizes gradients.
\sys therefore treats policy dissemination as publish--subscribe over immutable versioned artifacts, where failures, joins, and leaves all become subscriber events recoverable by replay rather than consensus across survivors.
This yields \textit{non-blocking elasticity}: Actors and regions join or leave without pausing survivors. The one RL-specific complication, version-coupling within the one-step lag, is resolved by the design choices below.

\para{Per-prompt leases, not collective micro-batches.}
Prior systems divide work into micro-batches that participate in a synchronous gradient aggregation: losing a worker invalidates its micro-batches, forcing either a checkpoint-restored redo~\cite{wang2023gemini} or a peer reroute backed by replica state~\cite{jang2023oobleck,thorpe2023bamboo,gandhi2024recycle}.
In \sys, work units are per-prompt leases (2--3$\times$ median completion time), and the Trainer aggregates rollouts stochastically rather than collectively.
When an Actor fails or partitions, only its claimed prompts expire back to the pool; the Trainer proceeds with whatever arrives.
This reframes failure from a correctness problem (recover an invalidated collective) to a throughput problem (replenish lost producers), which the scheduler already solves.
A single acceptance predicate at settlement keeps stale or corrupted rollouts out of training: the Trainer admits result $r$ for job $j$ iff $t_r \le t^{\text{expire}}_j$, $v_r = v_j$, and $h_r = h(v_j)$.
It is the only correctness mechanism in the data plane; everything else optimizes throughput.

\para{Version-predicate join and delta catch-up.}
Prior systems treat cluster membership as binary and recover rejoining workers by restoring full model state: a worker is either in the collective or requires reconfiguration (template switching~\cite{jang2023oobleck}, live pipeline repartitioning~\cite{duan2024parcae}), and catch-up always costs $O(|\text{model}|)$ from CPU memory~\cite{wang2023gemini}, a redundant replica~\cite{jang2023oobleck,thorpe2023bamboo}, or a peer-to-peer copy~\cite{gandhi2024recycle,duan2024parcae}, regardless of how far behind the worker is.
\sys makes both soft. Membership is a version predicate: Actor $a$ is eligible at step $v$ iff it holds version $v$ or $v{-}1$ (\autoref{line:eligible}), and catch-up replays only the missed deltas, each sparse at 1--3\% of weights. A joining Actor downloads the latest full-weight snapshot from the Checkpoint Store (a by-product of periodic checkpointing) and registers at version $v{-}1$; at the next scheduling step it enters the eligible pool and is treated identically to any lagging Actor, receiving \textsc{Commit}$(v)$ before dispatch (\autoref{line:commit}). An Actor that has fallen $K$ steps behind replays only the chain $\mathcal{D}_{v_a+1}, \ldots, \mathcal{D}_{v}$ through the staged-activation protocol (\autoref{sec:overlapped-transfer}), at cost $O(K \cdot |\mathcal{D}|)$; for Qwen3-8B, $K \times 202$~MB rather than 15.6~GB flat, so re-synchronization scales with \textit{missed updates} rather than model size. An Actor that slips outside the window is quietly excluded with decay $\alpha$ applied to its capacity estimate (\autoref{line:exclude}) and rejoins when caught up. Because the proportional capacity $T$ in \autoref{alg:scheduling} is recomputed every step, surviving Actors need no notification of joins or leaves: the scheduler absorbs expansion and contraction implicitly.

\para{Region elasticity at larger granularity.}
A region (a Relay plus its attached Actors) is itself a subscriber in the publish--subscribe model: the Relay caches the current policy version and forwards deltas to local Actors.
Adding a region introduces exactly one new cross-region transfer path regardless of Actor count, because the two-tier fanout collapses per-region cross-region traffic to $O(1)$ (\autoref{sec:overlapped-transfer}).
The new Relay bootstraps from the Checkpoint Store, is inserted into the Trainer's fanout tree at version $v{+}1$, and its Actors follow the per-Actor on-ramp.
Region removal, whether graceful or by failure, eliminates the single relay path; its Actors follow the standard leave path, and no surviving Actor is paused.
A Relay failure within a live region is not a coordination event either: affected Actors fall back to direct fetch from the Trainer or a peer Relay, continuing to produce rollouts on their current version until a new delta arrives.

\para{Spot instances as a concrete application.}
The spot deployment from \autoref{sec:opportunities} is the canonical hard case for routine fault tolerance.
\sys handles them without any additional machinery: cloud preemption notices invoke the graceful-leave path within the warning window; involuntary preemptions vacate their leases back to the pool within one lease duration; and a returning Actor re-enters the eligible pool after replaying $O(K \cdot |\mathcal{D}|)$ of catch-up deltas, a few hundred MB for minute-scale outages.
An all-spot Actor pool is therefore a direct deployment consequence of the mechanisms above, not a separate design.
Trainer-side faults remain orthogonal and are handled by standard checkpoint-and-restart, within which the one-step lag bound keeps in-flight Actor valid.

\subsection{Discussion and Future Work}
\label{sec:discuss}

\para{Geo-distributed training integration.} Existing geo-distributed training distrubtes trainer across regions and aggregate gradient, which is well studied~\cite{mastosdi2024,athlur2022varuna,chen2025crosspipe,lim2024accelerating}. Combining it with \sys's policy-transfer pipeline is natural extension and we leave it as future work.

\para{Super-large MoE support.} Super-large MoE models such as DeepSeek-V3~\cite{liu2024deepseek} (671B total, 37B activated per token) and Kimi-K2~\cite{kimi2025k2} should benefit even more from our design: only routed experts are updated per step, on top of \sys's element-wise sparsity. However, the GPU resources required for MoE evaluation is beyond our capability.

\section{Experiment}
\label{sec:eval}

\noindent
We build \sys ($\sim$5K lines of Python codes) atop the PrimeRL framework with PyTorch FSDP2~\cite{pytorch-fsdp2} and vLLM~\cite{kwon2023efficient} as underlying training and inference engines.

\subsection{Experimental Setup}
\label{sec:eval:setup}

\para{Testbed.}
We evaluate \sys across geo-distributed GPU deployments spanning multiple continents, running Ubuntu~24.04.1 with CUDA~12.6.
The Trainer uses FSDP2 for distributed optimization, while Rollout Actors use vLLM for inference.
By default, the Trainer is deployed in the U.S.\ and Actors in Canada, connected by a commodity cross-cloud WAN link whose measured bandwidth fluctuates between 500\,Mbps and 1\,Gbps, exhibiting the temporal and cross-provider variability commonly observed in cross-cloud connectivity.
We evaluate Qwen3 models~\cite{qwen3technicalreport} at 4B/8B/14B with 2/4/6$\times$H100 (SXM5) for training and 4/8/12 Actors, each on a single A100 (PCIe) GPU.
Unless otherwise stated, all experiments use this native cross-cloud WAN link.
To stress-test robustness, we additionally vary (i) bandwidth via \texttt{tc}~\cite{iproute2_tc} from 0.25 to 10~Gbps (\autoref{sec:eval:bandwidth}), (ii) geographic spread, with Actors placed in datacenters in Canada, Japan, the Netherlands, Iceland, and Australia (\autoref{sec:eval:multidc}), and (iii) inference heterogeneity, mixing A100 and L40 GPUs across Actors (\autoref{sec:eval:hetero}).

\para{Models and datasets.}
We use the GRPO algorithm~\cite{shao2024deepseekmath} on three reasoning benchmarks: Hendrycks MATH~\cite{hendrycksmath2021}, GSM8K~\cite{cobbe2021gsm8k}, and DeepScaleR~\cite{deepscaler2025}.
Each Actor generates a rollout group of size $B=512$.
Following recent RL systems~\cite{zhong2025streamrl,noukhovitch2025async,fu2025areal}, we adopt a one-step asynchronous policy: this bounded staleness preserves training quality while removing weight transfer from the critical path.
We run hundreds of optimizer steps per configuration and report throughput averaged after warmup.

\para{Baselines.}
We compare \sys against five baselines, all sharing the same Trainer implementation, Actor count, and configuration.
\textit{Ideal-SingleDC} colocates Trainer and Actors in a single datacenter over 800~Gbps RDMA and 900~GB/s NVLink; we construct it by replacing only the cross-cloud weight transfer cost in \sys's measured execution traces with the corresponding RDMA transfer cost, keeping training and rollout execution unchanged, yielding an idealized upper bound.
\textit{PrimeRL-Full} ports PrimeRL~\cite{primeintellect2025prime-rl} to a geo-distributed setting by broadcasting full model parameters every optimizer step under a one-step async bound (matching \sys for parity).
\textit{PrimeRL-Async2} and \textit{PrimeRL-Async4} relax this bound to 2 and 4 lag steps so Actors can overlap more of the dense broadcast with stale-weight rollouts, reflecting how PrimeRL would more aggressively be tuned for geo-distributed deployment.
\textit{PrimeRL-MultiStream} parallelizes broadcast over multiple TCP streams.

\para{Metrics.}
Following prior RL system work~\cite{zhong2025optimizing,zhong2025streamrl}, we report three metrics: \textit{step time} (wall-clock per optimizer step), \textit{throughput} (tokens/s system-wide), and \textit{accuracy} (per-step correct-answer reward on training problems).

\begin{figure*}[!t]
    \centering
    \includegraphics[width=\textwidth]{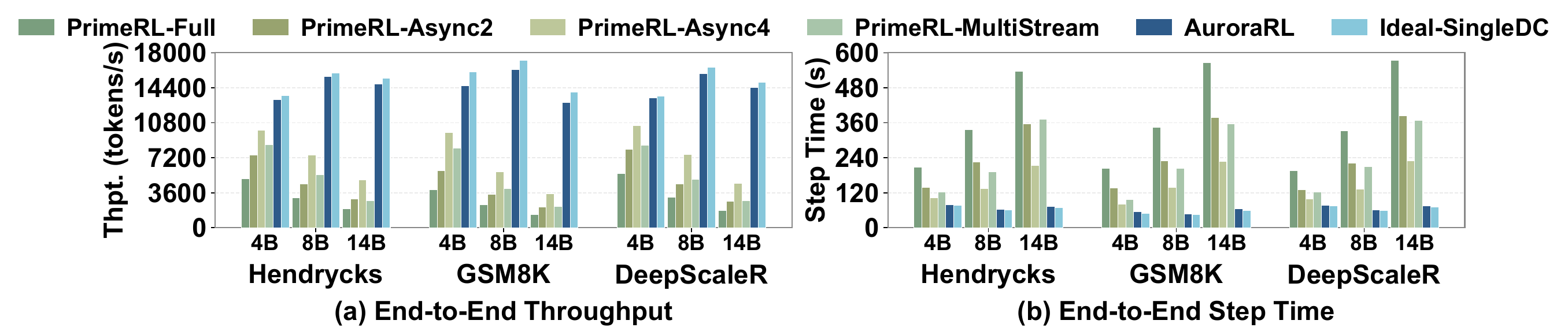}
    \caption{End-to-end (a) throughput and (b) step time across three benchmarks and Qwen3 model sizes. \sys approaches Ideal-SingleDC and outperforms all PrimeRL variants Full, Async2/4, MultiStream).}
    \label{fig:main_comparison}
\end{figure*}

\subsection{End-to-End Results}
\label{sec:eval:e2e}

\noindent
We first evaluate end-to-end RL training performance under real-world geo-distributed conditions.

\para{Throughput.}
\autoref{fig:main_comparison}(a) reports end-to-end throughput across all benchmarks and model sizes.
\sys consistently outperforms all four PrimeRL variants while closely approaching Ideal-SingleDC, and the gap widens with model size: 2.4--3.7$\times$ over PrimeRL-Full on Qwen3-4B and 7.7--9.5$\times$ on Qwen3-14B.
Relaxing the lag bound (Async2/Async4) buys 44--160\% over PrimeRL-Full and still trails \sys by 1.3--6.1$\times$ since WAN broadcast exceeds the rollout-overlap window (Qwen3-14B: $\sim$500\,s broadcast vs $\sim$400\,s from a 4-step lag). PrimeRL-MultiStream's parallel TCP transfer gains up to 2.1$\times$ but still trails \sys by 1.5--6.0$\times$, since dense weight transfer remains the binding WAN cost.
In contrast, \sys keeps weight transfer off the critical path, so end-to-end performance is governed by rollout generation and training computation.

\para{Step time.}
\autoref{fig:main_comparison}(b) shows average step time.
PrimeRL-Full step time exceeds 500\,s for Qwen3-14B. Relaxing the lag bound (Async2/Async4) shaves 30--60\% step time, but dense delivery still remains on the critical path.
PrimeRL-MultiStream parallelizes the broadcast but the dense transfer remains on the critical path.
In contrast, \sys reaches step times close to Ideal-SingleDC by combining sparse delta updates (\autoref{sec:delta-checkpoints}) with streaming delta transfer (\autoref{sec:overlapped-transfer}), hiding synchronization latency off the critical path.

\para{Gap to ideal performance.}
\sys is within 1.31--8.91\% of Ideal-SingleDC throughput across all configurations; the small residual gap comes from WAN latency and CPU-side delta extraction (neither exists on RDMA), both largely overlapped with rollout. Overall, \sys narrows the gap from 59.0--90.3\% under PrimeRL-Full (37.3--84.5\% under PrimeRL-MultiStream) to within 1.31--8.91\%.

\begin{figure}[!t]
\centering
\includegraphics[width=0.4\textwidth]{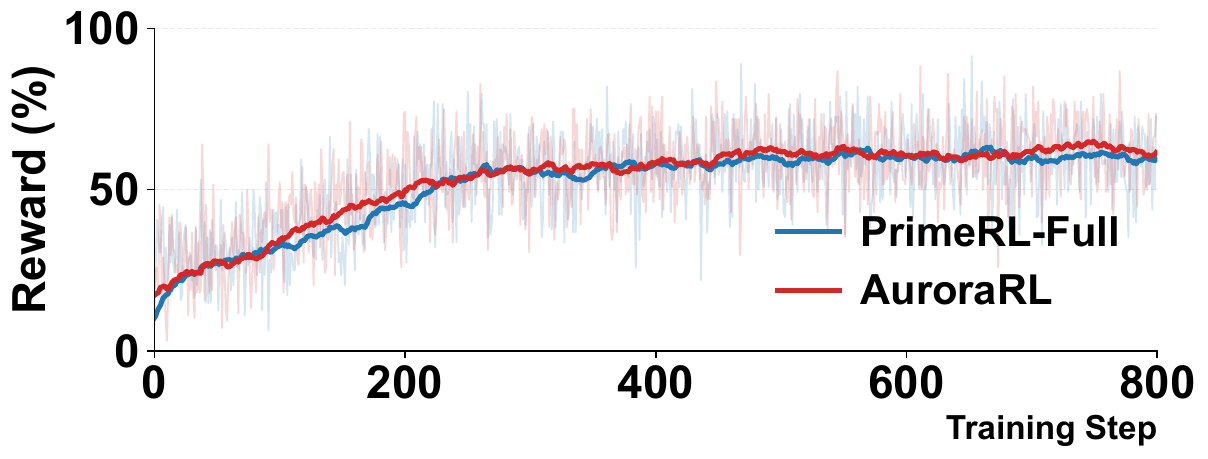}
\caption{Training reward curves of PrimeRL-Full vs.\ \sys on Qwen3-4B / DeepScaleR.}
\label{fig:reward}
\end{figure}

\para{Accuracy verification.}
We next verify \sys doesn't sacrifice model quality. We train Qwen3-4B on DeepScaleR~\cite{deepscaler2025} with both \sys and the baseline under identical hyperparameters.
\autoref{fig:reward} shows the two correct-answer reward curves track each other across 800 steps and both converge to $\sim$60\% mean reward in the final 100 steps, indicating that \sys's throughput gains come at no cost to model quality.

\begin{table}[t]
  \centering
  \small
  \caption{Downstream tasks accuracy evaluation.}
  \label{tab:task_eval}
  \begin{tabular}{lcccc}
  \toprule
  Benchmark & \sys & Full & Async2 & Async4 \\
  \midrule
  Math500 Pass@1   & 0.9060  & 0.9040  & 0.8880 & 0.8900 \\
  AIME2024 Pass@16 & 0.7267 & 0.7280 & 0.6740 & 0.6300 \\
  \bottomrule
  \end{tabular}
\end{table}

\para{Downstream tasks evaluation.}
To further assess task-level accuracy, we evaluate checkpoints after 800 DeepScaleR RL training steps on Math500~\cite{lightman2023lets} and AIME2024.
As shown in \autoref{tab:task_eval}, \sys closely matches PrimeRL-Full (0.906 vs.\ 0.904 on Math500 pass@1, 0.7267 vs.\ 0.7280 on AIME2024 pass@16), confirming that \sys's throughput gains come without quality degradation. PrimeRL-Async2 and PrimeRL-Async4 score lower on both benchmarks, suggesting that larger policy lagis the main source of quality degradation.

\begin{figure}[t]
\centering
\includegraphics[width=0.42\textwidth]{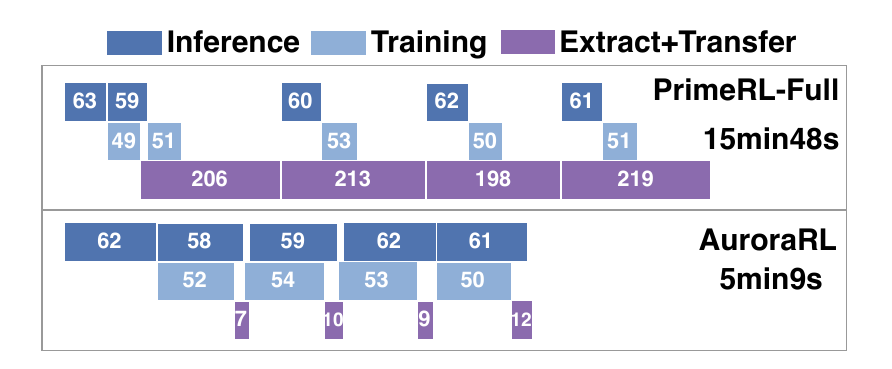}
\caption{Execution timeline for five Qwen3-8B steps: PrimeRL-Full vs \sys.}
\label{fig:timeline}
\end{figure}

\subsection{Ablation Study}
\label{sec:eval:abla}

\noindent
Then, we isolate each mechanism's contribution using Qwen3-8B with the same configuration as \autoref{sec:eval:e2e}.

\para{Execution timeline.}
We first validate that synchronization is removed from the critical path end-to-end.
\autoref{fig:timeline} traces five consecutive Qwen3-8B steps for PrimeRL-Full and \sys.
PrimeRL-Full broadcasts full weights every step ($\sim$200\,s), so even with one-step policy lag the dense transfer surfaces as per-step overhead and five steps take 15\,min~48\,s.
\sys completes the same five steps in 5\,min~9\,s, a $3.1\times$ speedup: sparse deltas shrink the per-step payload from 15.6\,GB to 202\,MB and the entire 7--12\,s of extraction-plus-transfer overlaps with rollout generation.

\begin{figure}[t]
\centering
\includegraphics[width=0.45\textwidth]{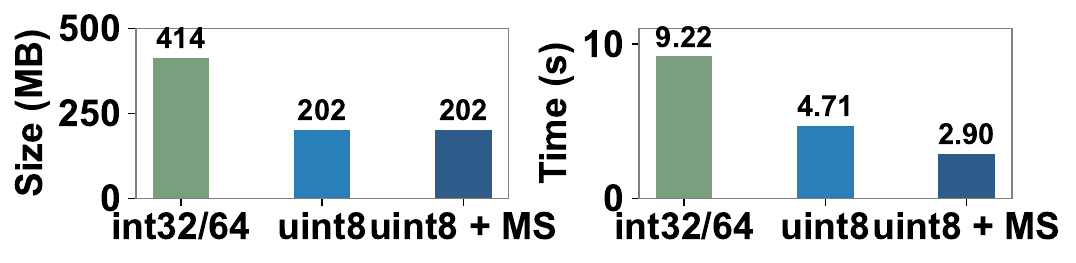}
\caption{Per-step delta transfer cost for Qwen3-8B (US--Canada). MS denotes multi-stream transport.}
\label{fig:encoding_multistream}
\end{figure}

\para{Encoding optimization.}
We next isolate how much each transfer-side mechanism contributes to per-step cost.
\autoref{fig:encoding_multistream} reports the per-step delta transfer cost for Qwen3-8B over the US--Canada WAN.
Compared to naive \texttt{int32/64} index encoding, the \texttt{uint8} delta encoding (\autoref{fig:delta_representation}) halves the payload from 414\,MB to 202\,MB and cuts transfer time from 9.22\,s to 4.71\,s; layering multi-stream on top further reduces it to 2.90\,s, a $3.2\times$ overall speedup from combining compact encoding with link saturation.

\begin{figure}[!t]
\centering
\includegraphics[width=0.45\textwidth]{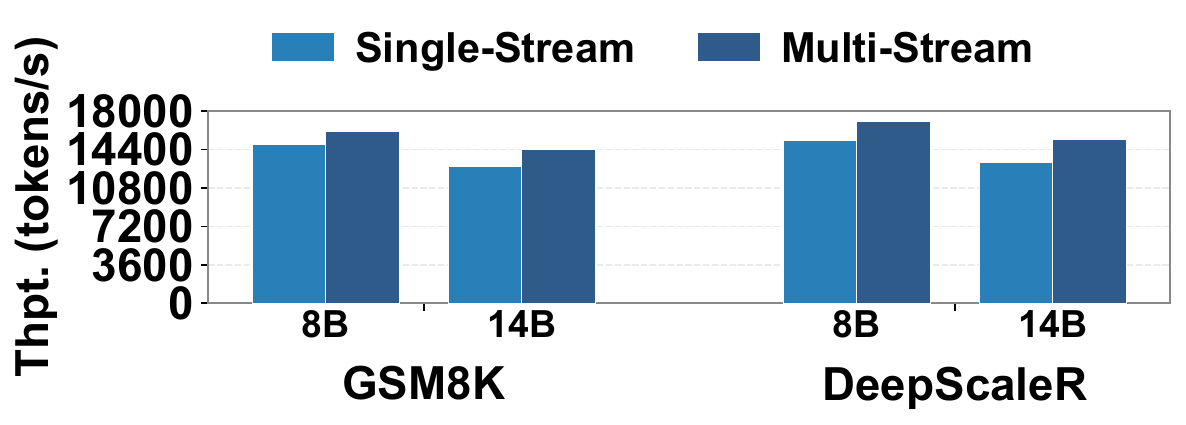}
\caption{Throughput with single-stream versus multi-stream delta transfer across two datasets.}
\label{fig:multistream_throughput}
\end{figure}

\para{Streaming optimization.}
Beyond per-transfer cost, we ask whether transport-level parallelism translates into end-to-end throughput gains once sparse deltas are in place.
\autoref{fig:multistream_throughput} shows that four parallel TCP streams improve throughput by 8.2\%/11.7\% on Qwen3-8B and 12.4\%/16.3\% on Qwen3-14B (GSM8K/DeepScaleR).
Larger models benefit more because their delta payloads occupy a bigger fraction of step time, so transport parallelism yields proportionally larger end-to-end gains in a pipeline where training and rollout generation already dominate the step budget.

\begin{table}[t]
  \centering
  \small
  \caption{End-to-end throughput with and without relay}
  \label{tab:relay_throughput}
  \begin{tabular}{lccc}
  \toprule
  Dataset & Baseline & Relay & Improvement \\
  \midrule
  GSM8K & 13588.4 & 14188.1 & +4.4\% \\
  DeepScaleR & 14060.0 & 16012.7 & +13.9\% \\
  \bottomrule
  \end{tabular}
\end{table}

\para{Relay optimization.}
When multiple Actors share a remote region, the Trainer's outbound WAN becomes the bottleneck; \sys then sends each delta once to a Relay that fans it out locally, cutting cross-region traffic to $1/N$.
\autoref{tab:relay_throughput} reports end-to-end throughput with and without Relays under a Canada--Australia deployment for Qwen3-8B: Relays improve throughput by 4.4\% (GSM8K) and 13.9\% (DeepScaleR), with the benefit expected to grow with model size.

\subsection{Heterogeneous GPU Results}
\label{sec:eval:hetero}

\noindent
Cross-cloud inference pools are typically heterogeneous in GPU type and link quality; we ask how much heterogeneity costs under a naive scheduler and how much our throughput-aware load balancer recovers.
Training Qwen3-4B on 4$\times$H100 in Canada with a mixed pool of 4$\times$A100 + 4$\times$L40 in the US, \autoref{tab:load_balancing_throughput} shows that uniform assignment leaves the faster A100s waiting on slower L40s, while throughput-proportional balancing improves end-to-end performance by 35.5\% on GSM8K and 26.4\% on DeepScaleR, confirming that heterogeneity-aware scheduling is essential for imbalanced Actors.

\begin{table}[t]
  \centering
  \small
    \caption{End-to-end throughput (tokens/s) under heterogeneous inference deployment (A100 + L40) with uniform versus heterogeneity-aware load balancing.}
  \label{tab:load_balancing_throughput}
  \begin{tabular}{lccc}
  \toprule
  Dataset & Uniform & Heterogeneity-aware & Improvement \\
  \midrule
  GSM8K & 10414.6 & 14107.6 & +35.5\% \\
  DeepScaleR & 10755.1 & 13589.9 & +26.4\% \\
  \bottomrule
  \end{tabular}
\end{table}

\subsection{Fault Tolerance}
\label{sec:eval:fault}

\begin{figure}[!t]
\centering
\includegraphics[width=0.4\textwidth]{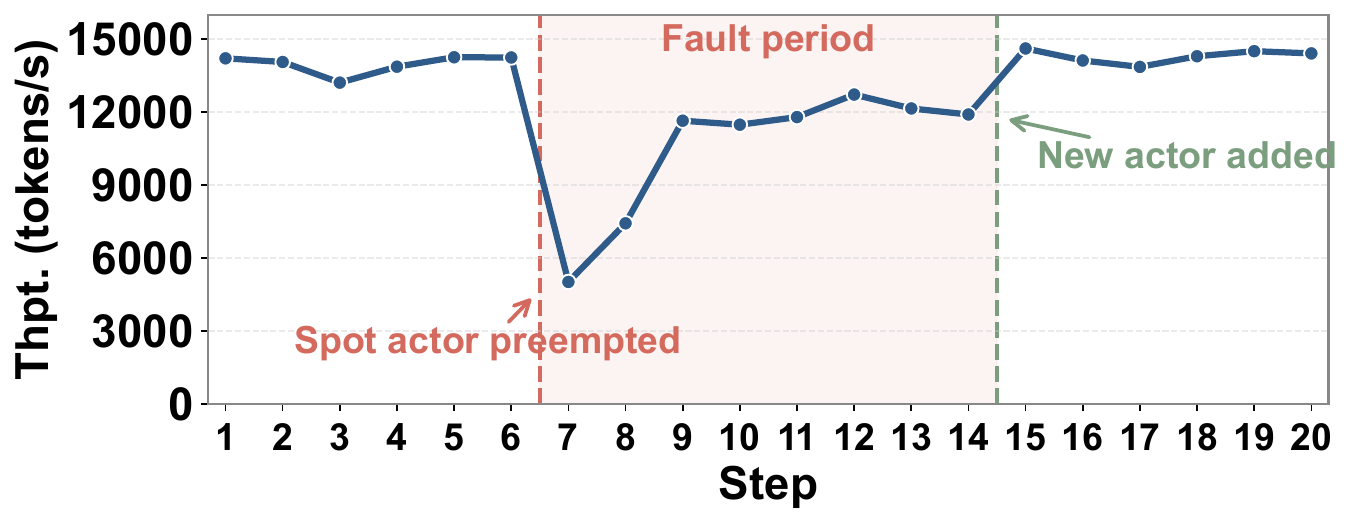}
\caption{Fault tolerance (Qwen3-4B, 4 Actors): one Actor preempted at step 6, replacement Actor onboarded at step 14.}
\label{fig:fault}
\end{figure}

\noindent
Cross-cloud deployments routinely lose Actors mid-run.
Since GPU faults, host outages, and other failure modes appear to the system identical to spot preemption (an Actor disappears and later returns), we use Google Cloud spot instances as the representative test: with Qwen3-4B and four spot-backed Actors, a preemption occurs at step 6 and we onboard a freshly leased Actor at step 14.
\autoref{fig:fault} shows throughput drops at step 6 from $\sim$13.4K to $\sim$4.7K tokens/s as the preempted Actor's rollout aborts; the remaining three Actors continue at $\sim$11--12K tokens/s without a global pause, then rapidly recover to $\sim$13.5K tokens/s after the new Actor's on-ramp at step 14, confirming that \sys handles partial failures and supports dynamic membership changes on unreliable cross-cloud capacity.

\begin{table*}[!t]
\centering
\caption{Cost efficiency of \sys vs.\ Ideal-SingleDC. Cross-cloud uses on-demand H100s from Hyperbolic~\cite{hyperbolic2025} and A100s from Prime Intellect~\cite{primeintellect2025} over standard VM networking; SingleDC uses reserved Hyperbolic RDMA clusters. Throughput is the geometric mean over Hendrycks MATH, GSM8K, and DeepScaleR; egress fees are negligible vs.\ GPU cost.
}
\label{tab:cost-efficiency}
\small
\resizebox{0.9\textwidth}{!}{
\begin{tabular}{l l l c c c c c}
\toprule
\textbf{Model} & \textbf{Method} & \textbf{Configuration} & \textbf{BW} &
\textbf{GM Throughput} & \textbf{\$/hr} & \textbf{tokens/\$} & \textbf{Norm.} \\
 &  &  &  & (tokens/s) &  & ($\times 10^{6}$) & (SingleDC=1) \\
\midrule
\multirow{2}{*}{Qwen3-8B}
 & \sys & 4$\times$H100 + 8$\times$A100 (cross-cloud on-demand) & $\sim$1\,Gbps & $\sim$15.9k & 15.88 & $\sim$3.60  & \textbf{1.21$\times$} \\
 & SingleDC & 1$\times$8$\times$H100 RDMA cluster (reserved) & $\ge$800\,Gbps & $\sim$16.5k & 19.92  & $\sim$2.99 & 1.0$\times$ \\
\midrule
\multirow{2}{*}{Qwen3-14B}
 & \sys & 6$\times$H100 + 12$\times$A100 (cross-cloud on-demand) & $\sim$1\,Gbps & $\sim$14.0k & 23.82 & $\sim$2.12  & \textbf{1.59$\times$} \\
 & SingleDC & 2$\times$8$\times$H100 RDMA cluster (reserved) & $\ge$800\,Gbps & $\sim$14.8k & 39.84  & $\sim$1.33 & 1.0$\times$ \\
\bottomrule
\end{tabular}
}
\end{table*}

\subsection{Sensitivity to Network Bandwidth}
\label{sec:eval:bandwidth}

\begin{figure}[!t]
\centering
\includegraphics[width=0.47\textwidth]{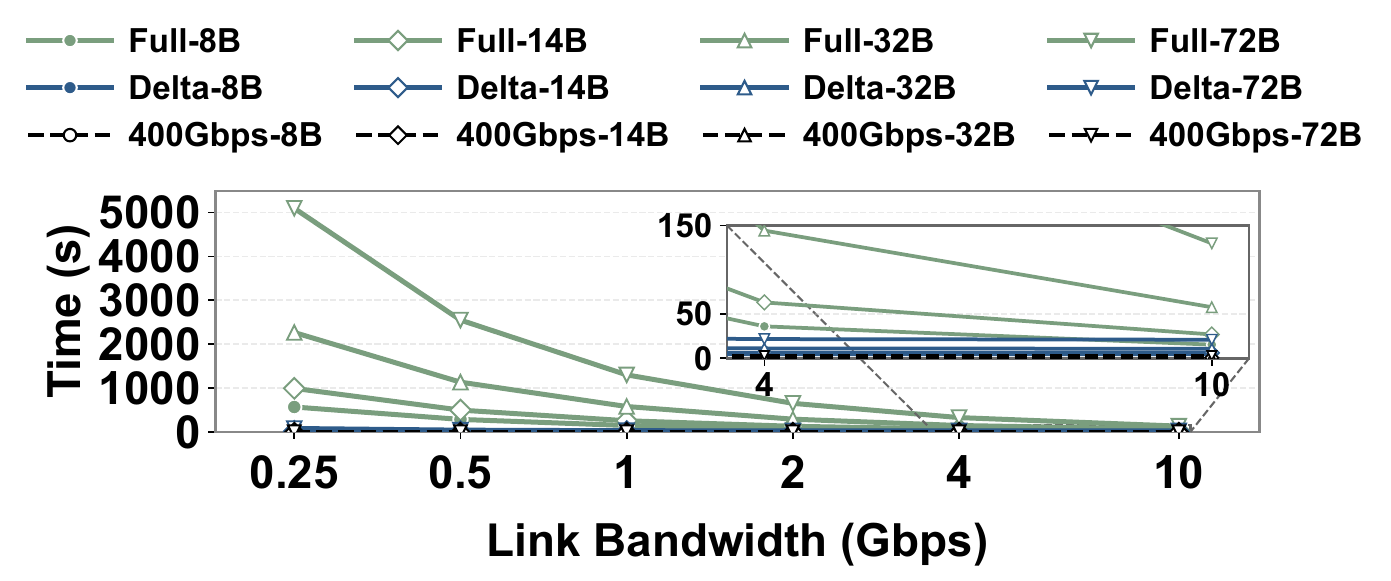}
\caption{Per-step weight transfer time under emulated bandwidth constraints. }
\label{fig:transfer_time_bandwidth_8b}
\end{figure}

\noindent
Beyond the default $\sim$1\,Gbps WAN, real cross-cloud bandwidth varies widely; we ask how each system's transfer cost behaves as the link tightens, sweeping emulated links from 250\,Mbps to 10\,Gbps using \texttt{tc}~\cite{iproute2_tc}.
In \autoref{fig:transfer_time_bandwidth_8b}, PrimeRL-Full (\textit{Full}) scales inversely with bandwidth, rising from 17.3\,s at 10\,Gbps to 566\,s at 250\,Mbps for Qwen3-8B and growing further with model size.
\sys (\textit{Delta}) is far less sensitive: at 10\,Gbps it transfers Qwen3-8B deltas in 0.25\,s, matching the 0.32\,s required by an ideal 400\,Gbps RDMA broadcast.
This holds across model scales, showing sparse delta transfer effectively decouples step time from raw link capacity.

\subsection{Multi-Datacenter Results}
\label{sec:eval:multidc}

\begin{figure}[!t]
    \centering
    \includegraphics[width=0.45\textwidth]{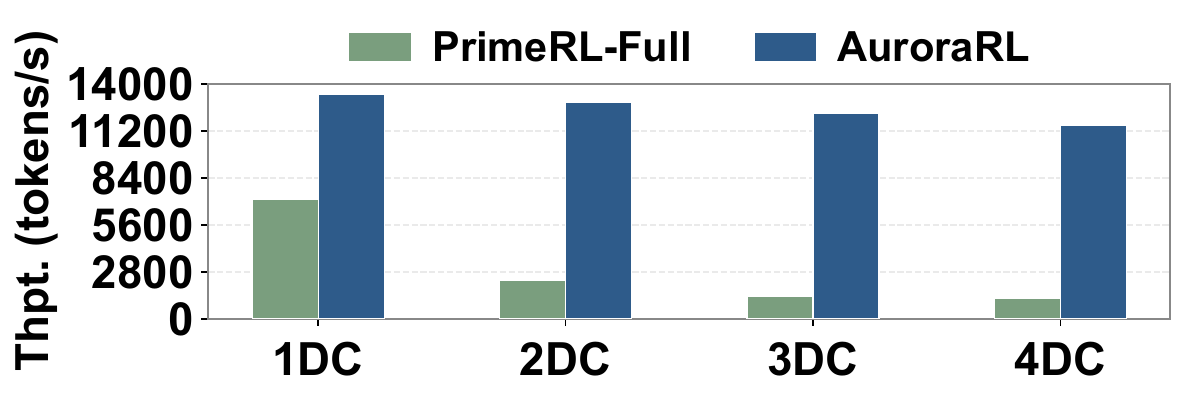}
    \caption{End-to-end throughput as Actors span 1--4 geographically distributed data centers (DCs) (Qwen3-4B, 2$\times$H100 Trainer, 4$\times$A100 Actors).}
    \label{fig:multidc_throughput}
\end{figure}

\noindent
Beyond the two-DC setup, \sys is designed to pool GPUs across many regions; we measure how throughput scales as inference Actors spread across continents.
The Trainer runs Qwen3-4B on 2$\times$H100 in the US, with four A100 Actors placed across one to four regions: Canada only (1-DC), Canada+Japan (2-DC), +Netherlands (3-DC), +Iceland (4-DC).
As shown in \autoref{fig:multidc_throughput}, PrimeRL-Full throughput collapses with added DCs, dropping from 7{,}137 to 1{,}219 tokens/s ($5.86\times$ reduction), since its broadcast is bounded by the slowest Actor.
\sys decreases by only 13.7\% from 1-DC to 4-DC, $1.9$--$9\times$ higher than PrimeRL-Full, showing sparse delta transfer scales gracefully across regions while dense broadcast does not.

\subsection{Practical Properties}
\label{sec:eval:practical}

\noindent
Finally, we evaluate scalability and cost-effectiveness.

\begin{figure}[!t]
\centering
\includegraphics[width=0.4\textwidth]{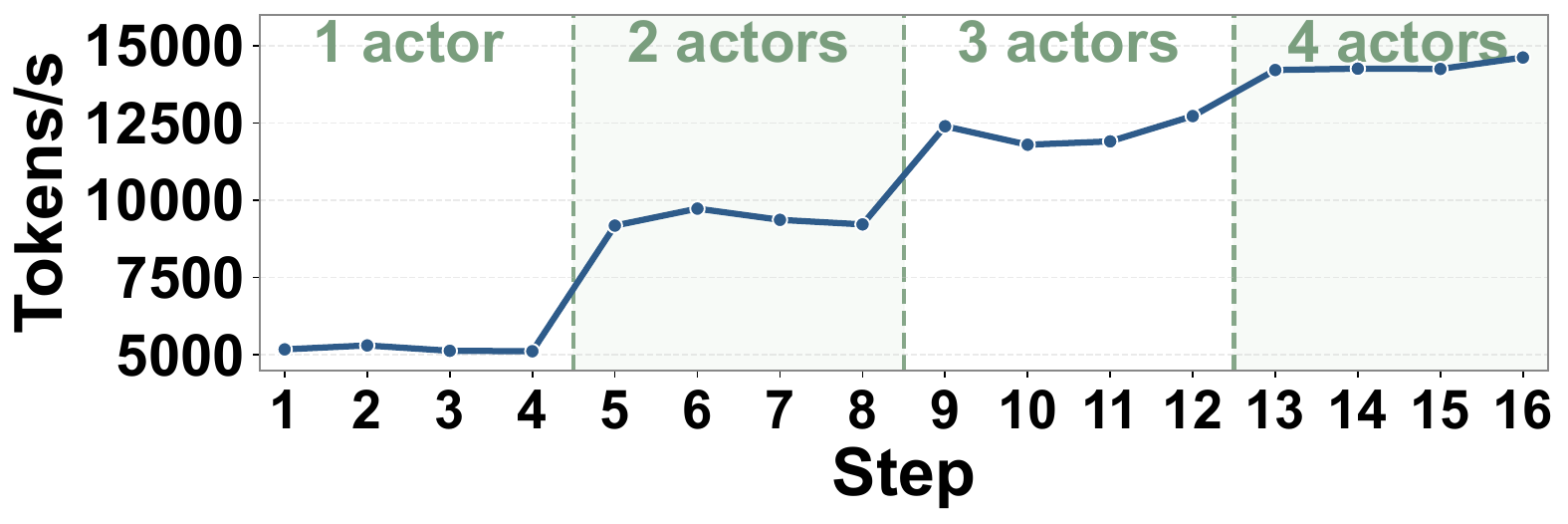}
\caption{Scalability (Qwen3-4B, batch 512): Actors increased from 1 to 4 over time.}
\label{fig:scale}
\end{figure}

\para{Scalability.}
A central promise of \sys is that operators can add opportunistic Actors and see proportional throughput; we test this by scaling the Actor pool from 1 to 4 mid-run on Qwen3-4B (batch 512).
\autoref{fig:scale} shows throughput grows from $\sim$5.6K to $\sim$9.3K, $\sim$11.5K, and $\sim$13.5K tokens/s as Actors scale 1$\to$2$\to$3$\to$4, a 2.4$\times$ total gain with diminishing returns, confirming that \sys absorbs additional Actors at runtime without restart or retuning.

\para{Cost-effectiveness analysis.}
To assess economic viability, we measure tokens per dollar against Ideal Single-DC at representative prices (\autoref{tab:cost-efficiency}). RDMA clusters come in fixed 8-GPU H100 blocks; with 1$\times$H100 $\approx$ 2$\times$A100 on our workload, 4$\times$H100+8$\times$A100 $\approx$ 1$\times$8$\times$H100 and 6$\times$H100+12$\times$A100 rounds up to 2$\times$8$\times$H100.
Even amortizing RDMA's multi-day minimums to hourly rates (which favors Single-DC), \sys reaches throughput within $\sim$5\% at substantially lower hourly cost, improving tokens/\$ by 1.21$\times$ (Qwen3-8B) and 1.59$\times$ (Qwen3-14B), demonstrating measurable cost-efficiency gains against RDMA baseline.

\section{Related Work}
\label{sec:related}

\para{RL Systems in a Single Datacenter.}
As discussed in \autoref{sec:existing-systems}, state-of-the-art RL systems such as OpenRLHF~\cite{hu2024openrlhf}, veRL~\cite{sheng2025hybridflow}, and StreamRL~\cite{zhong2025streamrl} maximize throughput via hybrid execution and disaggregated architectures, and rollout accelerators including RhymeRL~\cite{he2025history}, APRIL~\cite{zhou2025april}, SortedRL~\cite{zhang2025sortedrl}, RollPacker~\cite{gao2025rollpacker}, RLHFuse~\cite{zhong2025optimizing}, and OPPO~\cite{yan2025oppo} reduce generation latency. These techniques assume high-bandwidth datacenter networks; \sys is complementary, integrating with them for intra-region efficiency while addressing the orthogonal inter-region weight-transfer problem.

\para{Cross-datacenter training.}
Geo-distributed training systems target \textit{training-time} communication: MAST~\cite{mastosdi2024} schedules across regions; Varuna~\cite{athlur2022varuna} and CrossPipe~\cite{chen2025crosspipe} use network-aware pipeline parallelism; StellaTrain~\cite{lim2024accelerating} combines compression with bounded staleness; Oobleck~\cite{jang2023oobleck} and Bagpipe~\cite{agarwal2023bagpipe} improve robustness and overlap. They optimize gradients among trainers, whereas cross-region RL faces a different critical path: frequent policy sync from a centralized optimizer to dispersed rollout workers.

\section{Conclusion}

\sys enables practical RL post-training of LLMs over geo-distributed GPUs without sacrificing accuracy.
By co-designing with RL weight-update sparsity (sparse deltas, pipelined multi-stream transfer, and heterogeneity-aware scheduling), \sys closes the WAN-vs-RDMA gap.

\balance
\raggedbottom
\bibliographystyle{plain}
\bibliography{ref}

\newpage
\appendix
\section{Quanization Numeric Analysis}
\label{sec:appendix}

We analyze how representing checkpoints in BF16 affects checkpoint-to-checkpoint model updates observed by inference actors. Since inference actors consume BF16-represented checkpoints, the relevant state transition is the difference between consecutive BF16 checkpoints. For this analysis, we additionally save FP32 weights for a set of consecutive checkpoints offline. These FP32 checkpoints are used only to analyze the relationship between one-step FP32 updates and actor-visible BF16 checkpoint deltas

Let $W_t^{32}$ and $W_{t+1}^{32}$ denote the FP32 weights at two consecutive training steps, and let
\begin{equation}
\Delta_t^{32} = W_{t+1}^{32} - W_t^{32}.
\end{equation}
We consider the BF16 checkpoint delta
\begin{equation}
\Delta_{t,\mathrm{ckpt}}^{16} = \mathrm{bf16}(W_{t+1}^{32}) - \mathrm{bf16}(W_t^{32}),
\end{equation}
which directly captures the checkpoint-to-checkpoint state change observed by inference actors.

For normalized values, the spacing between adjacent BF16 numbers is approximately
\begin{equation}
\mathrm{ULP}_{\mathrm{bf16}}(W_i) \approx |W_i| \cdot 2^{-7}.
\end{equation}
Tables~\ref{tab:qwen_weight_distribution} and~\ref{tab:qwen_delta_distribution} provide coarse intuition for the scale mismatch between weight magnitudes and one-step FP32 updates in Qwen3-8B. Most weights lie in the range $10^{-3}$ to $10^{-1}$, whereas most one-step updates lie in the range $10^{-7}$ to $10^{-6}$.

\begin{table}[t]
\centering
\small
\caption{Distribution of weight magnitudes $|W|$ for Qwen3-8B.}
\label{tab:qwen_weight_distribution}
\begin{tabular}{lc}
\toprule
Range of $|W|$ & Fraction \\
\midrule
$0$ & 0.001\% \\
$(0, 10^{-4}]$ & 0.738\% \\
$(10^{-4}, 10^{-3}]$ & 4.232\% \\
$(10^{-3}, 10^{-2}]$ & 28.183\% \\
$(10^{-2}, 10^{-1}]$ & 66.750\% \\
$(10^{-1}, \infty)$ & 0.097\% \\
\bottomrule
\end{tabular}
\end{table}

\begin{table}[t]
\centering
\small
\caption{Distribution of one-step FP32 update magnitudes $|\Delta W|$ for Qwen3-8B.}
\label{tab:qwen_delta_distribution}
\begin{tabular}{lc}
\toprule
Range of $|\Delta W|$ & Fraction \\
\midrule
$0$ & 14.937\% \\
$(0, 10^{-7}]$ & 2.488\% \\
$(10^{-7}, 10^{-6}]$ & 82.561\% \\
$(10^{-6}, \infty)$ & 0.014\% \\
\bottomrule
\end{tabular}
\end{table}

These marginal distributions suggest that many one-step FP32 updates are small relative to the BF16 resolution at the corresponding weight scale. To quantify this effect element-wise, we compare each FP32 update to the local BF16 spacing:
\begin{equation}
r_i = \frac{|\Delta^{32}_{t,i}|}{\mathrm{ULP}_{\mathrm{bf16}}(W^{32}_{t,i})}.
\end{equation}
This ratio measures update magnitude relative to the local BF16 resolution. When $r_i < 1$, the update is smaller than one local BF16 spacing and is therefore likely to be invisible to actors. When $r_i > 1$, the update exceeds the local BF16 spacing and is thus more likely to remain visible to actors. In particular, values of $r_i < 0.5$ indicate that the update is far below the local BF16 resolution.

Empirically, the delta between consecutive BF16 checkpoints is highly sparse: only 1.27\% of parameters with nonzero one-step FP32 updates remain nonzero after BF16 checkpoint differencing. Among the rest, virtually all satisfy $r_i < 0.5$. Thus, these updates are far below the local BF16 resolution at the corresponding weight scale and therefore do not change the actor state.

Overall, these results show that BF16 substantially sparsifies checkpoint-to-checkpoint updates: for most parameters, one-step FP32 changes are too small to have a visible effect at actors. As a result, the delta between consecutive BF16 checkpoints is highly sparse, which directly motivates encoding and transmitting BF16 deltas in \sys.

\end{document}